# Evidence for Interlayer Electronic Coupling in Multilayer Epitaxial Graphene from Polarization Dependent Coherently Controlled Photocurrent Generation

Dong Sun[1], Julien Rioux[2], J. E. Sipe[2], Yang Zou[1,4], Momchil Mihnev[1], Claire Berger[3], Walt A. de Heer[3], Phillip N. First[3], and Theodore B. Norris[1]

[1]Center for Ultrafast Optical Science, University of Michigan,

Ann Arbor, MI 48109-2099

[2]Department of Physics, University of Toronto, Ontario, Canada M5S 1A7

[3]School of Physics, Georgia Institute of Technology, Atlanta, GA 30332

[4]Key Laboratory of Quantum Information, University of Science and Technology of China,

Hefei, China 230026

Abstract:

Most experimental studies to date of multilayer epitaxial graphene on C-face SiC have indicated that the electronic states of different layers are decoupled as a consequence of rotational stacking. We have measured the third order nonlinear tensor in epitaxial graphene as a novel approach to probe interlayer electronic coupling, by studying THz emission from coherently controlled photocurrents as a function of the optical pump and THz beam polarizations. We find that the polarization dependence of the coherently controlled THz emission expected from perfectly uncoupled layers, i.e. a single graphene sheet, is not observed. We hypothesize that the observed angular dependence arises from weak coupling between the layers; a model calculation of the angular dependence treating the multilayer structure as a stack of independent bilayers with variable interlayer coupling qualitatively reproduces the polarization dependence, providing evidence for coupling.

Although single-layer graphene is of great interest due to its unique electronic properties, for many applications in both transport and optoelectronics, it is highly desirable to use many layers while maintaining the unique properties of single-layer graphene[1-6] In particular, multilayer epitaxial graphene (MEG) grown by thermal decomposition on SiC substrates and patterned via standard lithographic procedures has been proposed as a platform for carbon-based nanoelectronics and molecular electronics [1-3, 7, 8]. A variety of initial studies showed rotational stacking order in multilayer epitaxial graphene (in contrast to A-B stacking in graphite), leading to decoupling of the layers and a linear band structure just as in single-layer graphene [9-13]. Recent experiments have indicated that A-B stacked bilayers may be present in large multilayer stacks, but they constitute at most 10% of the layers [14]. Angle-resolved photoemission experiments have provided strong evidence that the band structure even for small rotation angles between adjacent layers remains identical to that of isolated graphene[15]. In this work, we describe an optical probe that is in principle very sensitive to interlayer coupling effects, and has the advantage that it is sensitive to all the layers in the sample, and not just the top few layers[15, 16].

We have recently reported a non-contact all-optical femtosecond coherent control scheme to inject ballistic electrical currents in MEG[17]. In this scheme (Fig. 1(a)), quantum interference between single-photon and two-photon absorption breaks the material symmetry and the photoinjected carriers are generated with an anisotropic distribution in k-space, giving rise to a net current which is detected via an emitted THz signal[18]. The current density generation rate associated with interference between single- and two-photon absorption processes of beams at $2\omega$ and $\omega$ is of the form: $\vec{J} = 2|\vec{\vec{\eta}}| : \vec{E}^{\omega} \vec{E}^{\omega} \vec{E}^{2\omega} \sin(2\phi_{\omega} - \phi_{2\omega})$, where $\vec{E}^{\omega, 2\omega}$ and $\phi_{\omega, 2\omega}$ are the optical fields and phases, and $\vec{\vec{\eta}}$ is a fourth rank current injection tensor whose symmetry properties are governed by the illuminated material. In any 2-dimensional isotropic medium like graphene and graphite, there are only two independent tensor elements; if we define the optical linear polarization axes as the X and Y directions, then the X and Y components of the generated THz field are as follows [19, 20]:

$$\begin{bmatrix} E_X^{THz} \\ E_Y^{THz} \end{bmatrix} \propto \begin{bmatrix} \eta_{xxxx} \cos^2 \theta + \eta_{xyyx} \sin^2 \theta \\ \dfrac{1}{2}(\eta_{xxxx} - \eta_{xyyx}) \sin 2\theta \end{bmatrix}$$

where $\theta$ is the polarization angle between the $\omega$ and $2\omega$ beams. This relationship can be further simplified by noting that $\eta_{xyyx} / \eta_{xxxx} = -1$ in single-layer graphene, as calculated in previous theoretical work [21]. For graphite, this relationship has been measured to be $\eta_{xyyx}/\eta_{xxxx}$=-0.19±0.03 for a fundamental beam $\omega$ at 1400 nm [18, 22]. In this paper, we measure the X and Y components of the THz field generated by coherently controlled photocurrents in MEG as a function of $\theta$, in order to determine whether the third-order nonlinear tensor in MEG is consistent with a model incorporating only isolated graphene layers.

The sample used is a MEG film produced on the C-terminated face of single-crystal 4H-SiC by thermal desorption of Si. Details of the growth process and sample characterization can be found in references [1-3]. The thickness of the sample is determined by ellipsometry and the samples used in our experiments are 30 and 62 layers respectively. The layer nearest the substrate is heavily doped ($10^{13}$ cm$^{-2}$), and the doping decreases rapidly to a density of order $10^9$ cm$^{-2}$ with a decay (screening) length of 1 layer [23]. For the experiments reported here, the small background doping density of the majority of the layers in the sample is irrelevant and does not affect the dynamics. The results from different samples are qualitatively the same.

As in our prior work demonstrating coherent control in MEG[24], a commercial 250 kHz Ti: sapphire oscillator/amplifier operating at 800 nm is used to pump an optical parametric amplifier (OPA), followed by a difference frequency generator (DFG) to generate pulses with average power of 2-3 mW at 3.2 μm ($\omega$ beam) and 200-fs pulse width. The $\omega$ beam passes through a AgGeS$_2$ crystal (type I) to generate the 2$\omega$ beam at 1.6 μm with P$^{2\omega}$=200 μW. The $\omega$ and 2$\omega$ pulses are separated into the two arms of a Michelson interferometer using a dichroic beamsplitter. The relative polarization between the two pulses is varied by rotating the $\omega$ beam polarization with a λ/2 waveplate in the $\omega$ arm; the relative phase between the $\omega$ and 2$\omega$ pulses is controlled using a piezoelectric optical delay stage in the $\omega$ arm. In all measurements, the polarization of the 2ω pulse is fixed and the polarization of the ω pulse is rotated by a λ/2 waveplate. The two pump beams emerging from the Michelson interferometer are overlapped on the samples with a 20-μm diameter (FWHM) spot size, producing peak focused intensities for the 3.2 μm and 1.6 μm beams of respectively 2.26 GW/cm$^{-2}$ and 0.32 GW/ cm$^{-2}$ on the sample, including losses due to all intermediate optics. The sample is held at room temperature.

The coherently injected photocurrent is detected via the emitted terahertz radiation in the far field by electro-optic sampling[25]. The sampling pulse is the remnant 800 nm beam from the OPA, compressed below 80 fs by a prism pair and co-propagating with variable delay through a 1-mm thick (110)-oriented ZnTe crystal. The THz-induced birefringence of the 800 nm probe beam is detected by a balanced detector that allows for a direct measurement of the emitted THz field. A polyethylene (HDPE) slab is inserted between the graphene sample and the ZnTe electro-optic crystal in order to block residual pump light passing through the sample, and pass only the emitted THz radiation. A wire-grid THz polarizer is inserted after the polyethylene (HDPE) slab to analyze the generated photocurrent direction. The wire-grid polarizer and ZnTe crystal are rotated together to detect THz radiation polarized either parallel or perpendicular to the 2ω pump beam. The effective bandwidth of the electro-optic detection system is estimated to be ~2 THz due to phase mismatch between the terahertz and probe beams.

A typical THz waveform generated from the coherently controlled photocurrent is shown in Fig 2(a). The oscillatory temporal waveform is a result of the finite bandwidth of the electro-optic detection system and water-vapor absorption rather than the dynamics in the sample. The THz peak field marked by the arrow in Figure 2(a) is well controlled by the relative phases of the $\omega$ and $2\omega$ beams through the phase parameter $\Delta\varphi=2\varphi_\omega-\varphi_{2\omega}$, as shown in Figure 2(b)[17]. $\Delta\varphi$ is tuned by the piezostage positioned in the $\omega$ arm of a Michelson interferometer. The reduction of the oscillation amplitude is due to the walk off between the $\omega$ and $2\omega$ pulses and thus the amplitude profile gives the cross correlation of $\omega$ and $2\omega$ beams of 250 fs. The peak THz field shows a sinusoidal dependence on $\Delta\varphi$ as predicted by theory. Due to possible phase drifts in the Michelson interferometer during a long measurement (arising from thermal, humidity and air flow fluctuations), a scan covering several cycles is repeated for each polarization angle and thus avoid the possibility of the spontaneous phase drift in Michelson during the measurement.

The experimental quantity of interest in this paper is the dependence of the THz amplitude on the relative polarizations of the $\omega$ and $2\omega$ pulses when the THz polarization is either parallel or perpendicular to the $2\omega$ beam polarization. We take the peak-to-peak value as shown in Fig. 2(b) as the THz amplitude, and plot the result as a function of polarization angle in Fig. 3. The inset of Fig. 3 shows the coordinate system used to describe the system: the $2\omega$ polarization is fixed along the X direction and $\theta$ is the angle between the $\omega$ and $2\omega$ beams. The red curve in the polar plot corresponds to THz radiation polarized along the X direction; the black curve as for THz along the Y direction. When the THz polarizer is aligned along Y (perpendicular to the $2\omega$ beam), the maximum THz emission occurs when the $\omega$ beam is polarized at $\theta=45^o$, $135^o$, $225^o$, and $315^o$.

The theoretically predicted[21, 26] polarization dependence of the THz emitted by coherently controlled photocurrents, under the same conditions as the experiment, is shown for single-layer graphene in Fig 4(a). Comparing the experiment with the predicted result for isolated graphene, we see that the y component for theory and experiment are at least in qualitative agreement, whereas there is a significant deviation for the X component. This indicates that for this probe, MEG cannot be considered to be completely uncoupled layers of graphene. We therefore consider the possible role of interlayer coupling on the angular dependence of the coherent control THz signal.

A complete theory of coherent control in MEG would need to incorporate the full wavefunctions of multilayer graphene with various non-Bernal rotations (i.e. rotations other than 60°). Because the rotation angles between adjacent layers can be small, commensurate rotations lead to very large supercells[9]. Aside from challenges of performing the calculation for large supercells, there remain significant discrepancies between experimental results such as ARPES and the predictions of tight-binding and *ab initio* calculations; for example, theory[27, 28] predicts a renormalization of the Fermi velocity for small angles which is not observed in experiments[15]. Hence, in order to begin to address whether interlayer electronic coupling could be responsible for our observed angular dependence of the coherent control, we apply a simpler model for the interlayer coupling: we assume that the effect on interlayer coupling will be similar to that of Bernal-stacked bilayers, where we take the interlayer coupling to be a parameter. There is some physical justification for such a model, since if there are coupled layers with small twist angles, there will be large regions of the sample which effectively have A-B alignment, and other regions which have A-A alignment. Indeed such local coupling has been observed in real-space mapping of magnetically quantized states in similar samples[16].

The bilayer response is characterized by the interlayer coupling energy $\gamma_1$ and a linewidth $\Gamma$ that arises because two-photon absorption can be resonant with an intermediate state[26]. The predicted result for the shape of the Y component is given by $\sin(2\theta)$, the same as for graphene and for any 2D isotropic medium. For $2\hbar\omega < \gamma_1$, the shape of the X component would also be the same as graphene. However, for $2\hbar\omega \approx \gamma_1$ or $2\hbar\omega \approx 2\gamma_1$, as it would be for the standard value for $\gamma_1 = 0.4$eV in graphite[29], the model predicts $\eta_{xyyx}/\eta_{xxxx} \approx -0.5$, reasonably independent of the value of $\Gamma$. This results in the angular distributions shown in Fig. 4(b). The predicted angular dependence is in qualitative agreement with experiment if we consider the limited angular accuracy. Of course, in the context of other experiments on the electronic structure of MEG, it is unlikely that MEG should be thought of as a stack of independent bilayers, but this model nonetheless shows that coupling between layers of graphene has a qualitative effect on the predicted coherently-controlled photocurrent. Figure 4(c) shows with a mixture of 70% single-layer and 30% bilayer with $\eta_{xyyx}/\eta_{xxxx} = 2.8$, the theoretical predicted angular dependence shows good agreement with that measured in our experiment.

In summary, we have demonstrated that pump pulse polarization dependent coherent controlled photocurrent measurement is a sensitive tool to observe interlayer coupling in multilayer epitaxial graphene. The observed polarization angular dependence differs notably from that expected for a single isolated graphene layer. A model calculation treating the electronic states as those of a bilayer with the interlayer coupling as a parameter qualitatively reproduces the observed angular dependence, thus indicating the presence of interlayer electronic coupling. Future work will focus on incorporating more realistic models of the electronic states in MEG.

We acknowledge financial support from NSF-MRSEC through contract DMR-0820382.

J.R. and J.E.S. acknowledge financial support from FQRNT and NSERC of Canada.

**Figure Captions:**

**FIG. 1** Schematic energy-momentum band diagram of doped and undoped layers of epitaxial graphene near the Dirac point. Red is associated with the $\omega$ beam, blue with the $2\omega$ beam. Asymmetric electron populations at $\pm k$ and hence current generation, is indicated by shaded patches. The dash line goes across the Fermi level of doped and undoped graphene layers.

**FIG. 2** (a) Time dependent electro-optic signals of THz fields emitted from coherently controlled photocurrent; (b) Terahertz field as function of the phase relationship between $2\omega$ and $\omega$ beam for constant time delay marked by the arrow in Fig a.;

**FIG. 3** Experimentally measured X (red) and Y (black) components of relative peak THz amplitude as a function of the polarization angle between $\omega$ and $2\omega$ pulse illustrated by the coordinates on the right corner on multilayer epitaxial graphene.

**FIG. 4** Theoretical predictions of current injection with linearly-polarized $\omega$ and $2\omega$ beams: Polar plots of the projections of $\dot{\mathbf{J}}$ parallel (X) and perpendicular (Y) to $\hat{\mathbf{e}}_{2\omega}$, as a function of the angle $\theta$ between the polarization vectors: (a) single-layer graphene ($\eta_{XYYX}/\eta_{XXXX}$=-1); (b) bilayer graphene using graphite's value of $\gamma_1$=0.4eV ($\eta_{XYYX}/\eta_{XXXX}$=-0.5); and (c) a mixture of 70% single-layer and 30% bilayer ($\eta_{XYYX}/\eta_{XXXX}$=2.8). The shade circles represent unit amplitude and dashed lines represent negative projections.

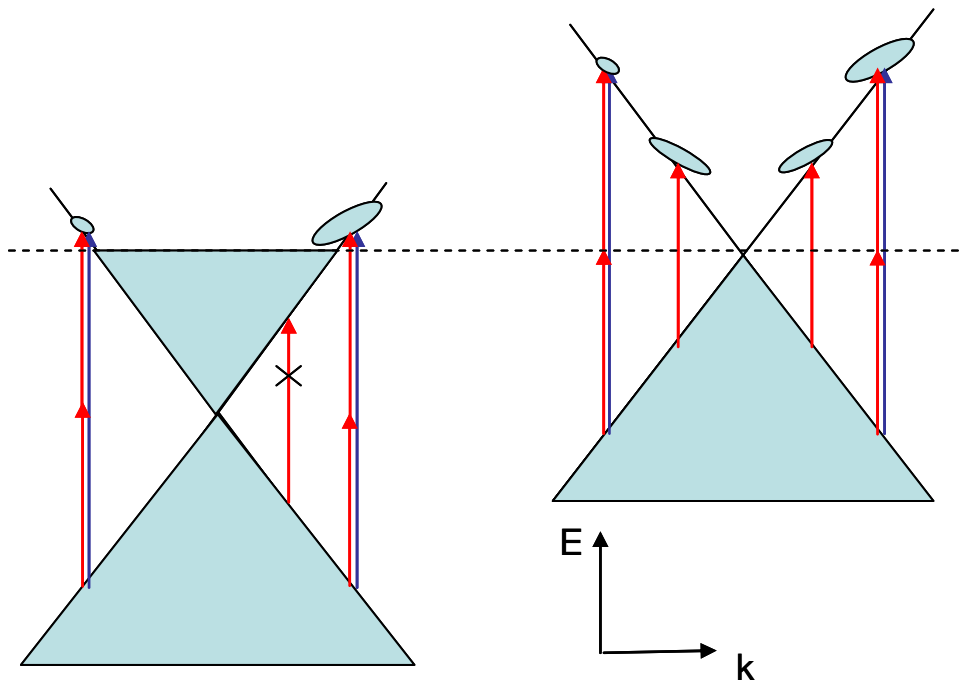

Figure 1

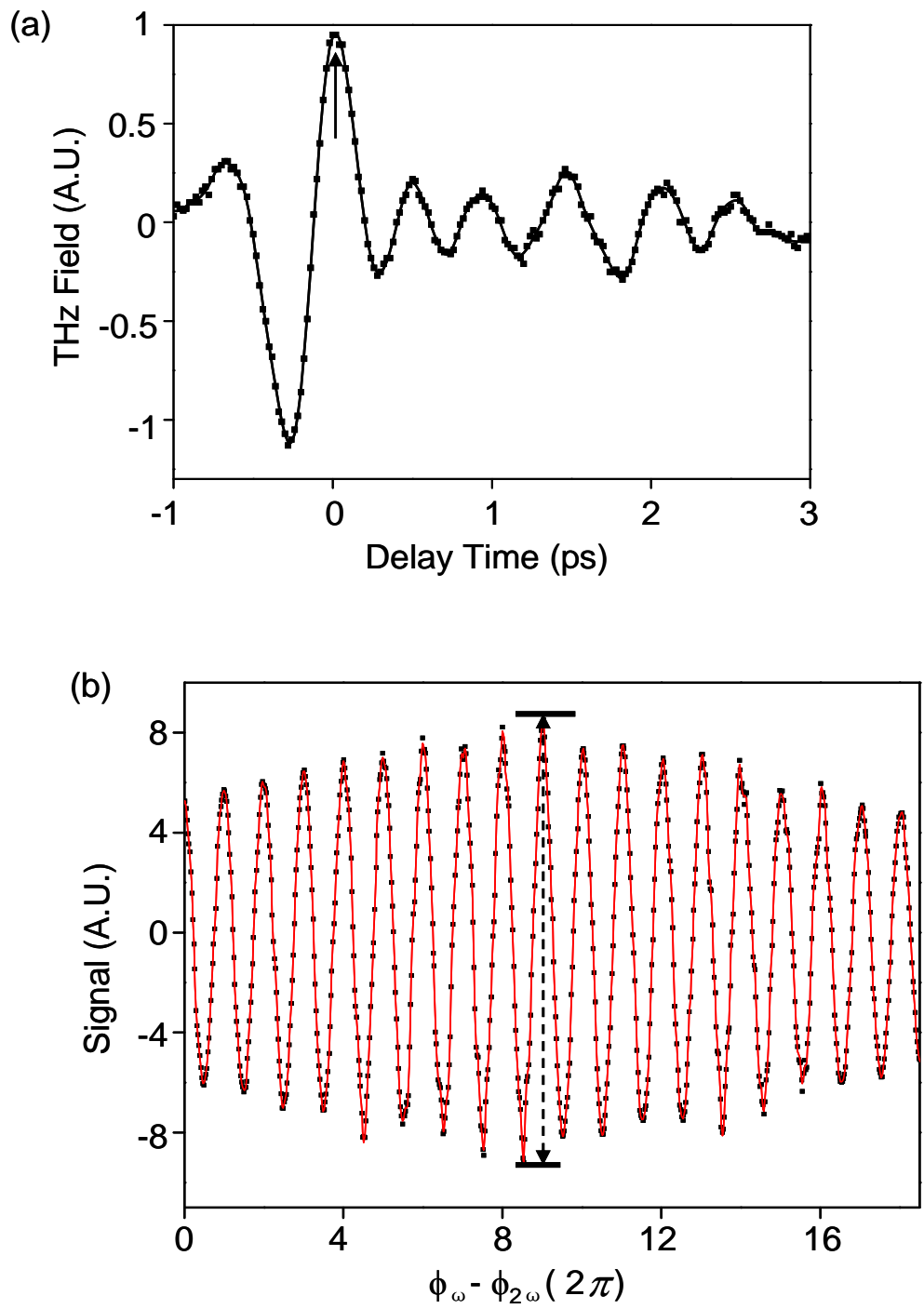

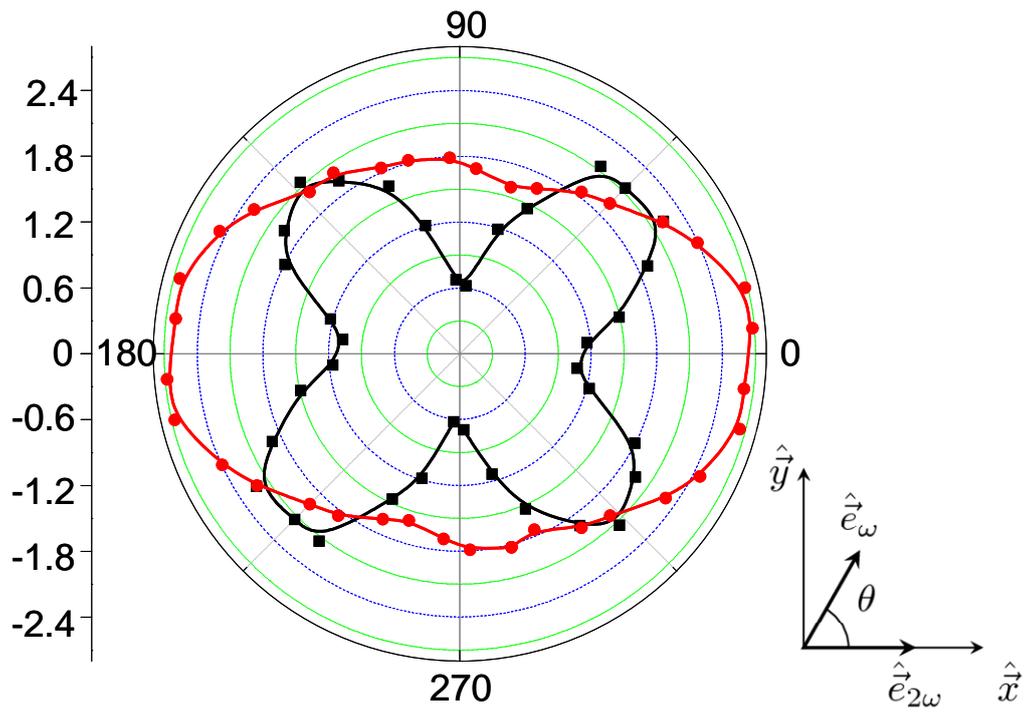

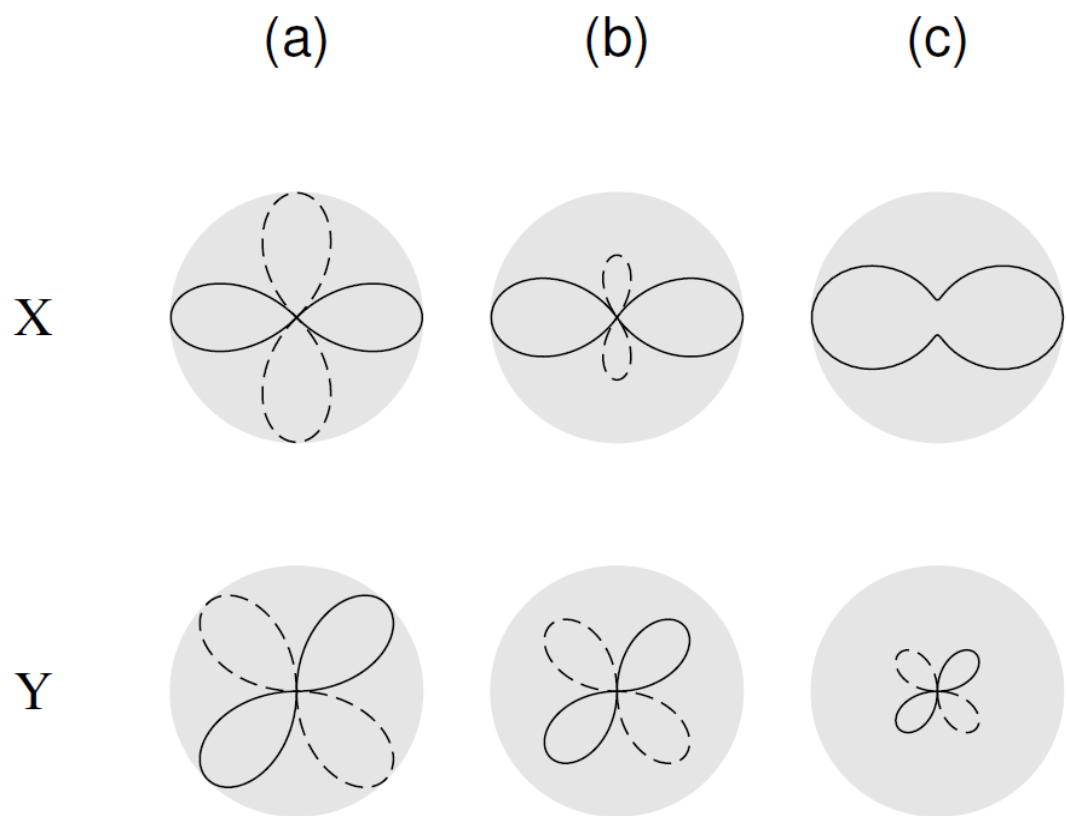

Figure 4